\documentstyle[12pt,epsfig]{article}
\makeatletter                    %allow @ in command names
\@addtoreset{equation}{section}  %reset equation count in each section
\makeatother                     %disallow @ in command names

\newskip\humongous \humongous=0pt plus 1000pt minus 1000pt

\newif\ifdtup

\def\be{\begin{equation}}
\def\ee{\end{equation}}

\begin{document}

\title{Explaining the Forward Interest Rate Term Structure}
\author{Andrew Matacz$^1$\thanks{E-mail: andrew.matacz@science-finance.fr} 
$~$and Jean-Philippe Bouchaud$^{1,2}$\thanks{Email: bouchaud@spec.saclay.cea.fr}\\
{\small $^1$ Science and Finance}\\
{\small 109-111 rue Victor Hugo} \\
{\small 92632 Levallois, France}\\
{\small http://www.science-finance.fr}\\
{\small $^2$ Service de Physique de l'Etat Condens\'e}\\
{\small CEA-Saclay, Orme des Merisiers} \\
{\small 91 191 Gif s/ Yvette, France}}
%\date{\small {\it Draft version -- Comments welcome --  July 1999}}
\maketitle

\begin{abstract}
We present compelling empirical evidence 
for a new interpretation of the Forward Rate Curve ({\sc frc}) term structure.
We find that the average {\sc frc} follows a square-root law, with 
a prefactor related to the spot volatility, suggesting a Value-at-Risk like 
pricing. We find a striking correlation between the instantaneous {\sc frc} and the 
past spot trend over a certain time horizon. This confirms the idea of 
an anticipated trend mechanism proposed earlier and provides a 
natural explanation for the observed shape of the {\sc frc} volatility. 
We find that the one-factor Gaussian Heath-Jarrow-Morton model calibrated to the 
empirical volatility function fails to adequately describe these features.

\end{abstract}

%\newpage
\section{Introduction}
The search for more adequate statistical models of the forward interest 
rate curve is essential for both risk control purposes and for a better 
pricing and hedging of interest rate derivative products \cite{Hull}. 
A large number 
of models have been proposed, but it is the Heath-Jarrow-Morton ({\sc hjm}) 
model \cite{hjm} that has become widely accepted as the most 
appropriate framework for addressing these issues. This model has been 
the basis for a large amount of research in relation to the pricing and hedging of 
derivative products. However comparatively little has addressed how well 
this model describes empirical properties of the 
forward rate curve ({\sc frc}).

In a previous paper \cite{FRC}, a series of observations concerning the 
U.S. {\sc frc} in the period 1991-96 were reported,
which were in disagreement with predictions of the standard models.
These observations motivated a new interpretation of {\sc frc} dynamics.
\begin{itemize}
\item First,
the average shape of the {\sc frc} is well fitted by a square-root law as
a function of maturity, with a prefactor very close to the spot rate volatility.
This strongly suggests that the forward rate curve is calculated by the money 
lenders using a {\it Value-at-Risk (VaR) like procedure}, and not, as assumed in standard 
models, through an averaging procedure. More
precisely, since the forward rate $f(t,\theta)$ is the agreed value at time $t$
of what will be the value of the spot rate at time $t+\theta$, a VaR-pricing 
amounts to writing:
\begin{equation}\label{VaR}
\int^{\infty}_{f(t,\theta)}dr'~P_M(r',t+\theta|r,t)=p,
\end{equation}
where $r(t)$ is the value of the spot rate at time $t$ and $P_M$ is the market 
implied probability of the future spot rate at time $t+\theta$. The value of 
$p$ is a constant describing the risk-averseness of money lenders. The risk is 
that the spot rate at time $t+\theta$, $r(t+\theta)$, turns out be larger than 
the agreed rate $f(t,\theta)$. This probability is equal to $p$ within the 
above VaR pricing procedure. If $r(t)$ performs a simple unbiased random walk, 
then Eq. (\ref{VaR}) indeed leads to 
$f(t,\theta)=r(t)+A(p)\sigma_r \sqrt{\theta}$, where $\sigma_r$ is the
spot rate volatility and $A(p)$ is some function of $p$.
\item Second, the volatility of the forward rate is found to be `humped' around 
$\theta=1$ year. This can be interpreted within the
above VaR pricing procedure as resulting from a time dependent {\it anticipated trend}. Within a VaR-like pricing, the {\sc frc}
is the envelope of the future anticipated evolution of the spot rate. 
On average, this evolution is unbiased, and the average {\sc frc} is a 
simple square-root. However, at any particular time $t$, the market  actually anticipates a future trend. It was argued in \cite{FRC} that this trend is
determined by the past historical 
trend of the spot rate itself over a certain time horizon. 
In other words, the market looks at the past and extrapolates the observed
trend in the future. This means that the probability distribution 
of the spot, $P_M(r',t+\theta|r,t)$, is not centered around $r$ but includes 
a maturity dependent bias whose magnitude depends on the historical 
spot trend. However, the market also knows that its estimate of the trend will not persist on the long run. The magnitude of this bias effect is expected to peak for a certain maturity and this can explain the volatility hump.
\end{itemize}

The aim of this paper is two fold. 
First we wish to empirically test the new interpretation 
of the {\sc frc} dynamics outlined above. Specifically, we 
report measurements over several different data-sets, of the shape of the 
average {\sc frc} and the correlation between the instantaneous 
{\sc frc} and the past spot trend over a certain time horizon.
We have investigated the empirical
behaviour of the {\sc frc} of four different 
currencies ({\sc usd}, {\sc dem}, {\sc gbp} and {\sc aud}),
in the period 1987-1999 for the {\sc usd} and 1994-1999 for the other 
currencies. Full report of the results can be found in \cite{frc2}. 
Here we only present detailed results for the {\sc usd} 
94-99, but we also discuss relevant results obtained with the other data-sets. 
Second, for {\sc usd} 94-99, we wish to compare these empirical results 
with the predictions of the one-factor Gaussian 
{\sc hjm} model fitted to the empirical volatility.

\section{Empirical Results}

Our study is based on data sets of daily 
prices of futures 
contracts on 3 month forward interest rates. In the {\sc usd} case 
the contract and exchange was the Eurodollar CME-IMM contract. 
In practice, the futures markets price three months forward rates for 
{\it fixed} expiration dates, separated by three month intervals. 
Identifying three months futures rates to 
instantaneous forward rates (the difference is not important here), 
we have available time series on forward rates
$f(t,T_i-t)$, where $T_i$ are fixed dates (March, June, September and 
December of each year), which we have converted into fixed maturity 
(multiple of three months) forward rates by a simple linear interpolation 
between the two nearest points such that 
$T_i - t \leq \theta \leq T_{i+1} - t$. In our notation we will identify 
$f(t,\theta)$ as the forward rate with fixed maturity $\theta$. 
This corresponds to the Musiela parameterization.
The shortest available maturity is $\theta_{\min}=3$ months, 
and we identify $f(t,\theta_{\min})$ to the spot rate $r(t)$. 
For the {\sc usd} 94-99 data-set discussed here, we had 38 maturities 
with the maximum maturity being 9.5 years.
We will define the `partial' spread $s(t,\theta)$, as the difference between 
the forward rate of maturity $\theta$ and the spot rate: 
$s(t,\theta)=f(t,\theta)-r(t)$. 
The theoretical time average of $O(t)$ will be 
denoted as $\langle O(t)\rangle$. We will refer to empirical averages (over a finite data set) as 
$\langle O(t)\rangle_e$. For infinite datasets the two averages are the same.

First we consider the average {\sc frc}, which can be obtained from 
empirical data by averaging the partial spread $s(t,\theta)$:
\begin{equation}
\langle s(t,\theta) \rangle_e = \langle f(t,\theta)-r(t) \rangle_e.
\end{equation}
In Figure 1 we show the average {\sc frc} $\langle s(t,\theta) \rangle_e$,
along with the following best fit:
\begin{equation}\label{eq35}
\langle s(t,\theta) \rangle_e=a\Bigl(\sqrt{\theta}-\sqrt{\theta_{\min}}\Bigr).
\end{equation}
As first noticed in \cite{FRC}, the
average curve can be quite satisfactorily fitted by a simple square-root law.
The corresponding value of $a$ (in $\%$ per $\sqrt{\mbox{day}}$) is 
0.049 which is very close to the daily spot volatility 0.047
(which we shall denote by $\sigma_r$). We have found precisely the same 
qualitative behaviour for our 12 year {\sc usd} data-set and also for the 
{\sc gbp} and {\sc aud}. The only exception was the steep {\sc dem} 
average {\sc frc} which can be explained by its low average spot 
level \cite{frc2}.
We have therefore greatly strengthened -- with much more empirical data -- 
the proposal of ref. \cite{FRC} that the {\sc frc} is on average fixed by a 
VaR-like procedure, specified by Eq. (\ref{VaR}) above.

In figure 2 we show the empirical 
volatility for the {\sc usd}, defined as:
\begin{equation}
\sigma(\theta)=\sqrt{\langle \Delta f^2(t,\theta)\rangle_e}, 
\qquad \sigma(\theta_{\min})\equiv \sigma_r,
\end{equation}
where $\Delta f(t,\theta)$ denotes the daily increment in the forward rates.
We see a strong peak in the volatility at 
1 year \cite{Hull,Moraleda,FRC}. For all the data-sets we have studied the 
volatility shows a steep initial {\it rise} between the spot rate and 6-9 
months forward \cite{frc2}. We also show the fit of the function:
\begin{equation}
\sigma(\theta)=0.061-0.014\exp\Big(-1.55(\theta-\theta_{\rm min})\Bigr)
+0.074~(\theta-\theta_{\rm min}) \exp\Bigl(-1.55(\theta-\theta_{\rm min})\Bigr).
\end{equation}
It is not {\it a priori} clear why the {\sc frc} volatility should 
{\it universally} be strongly increasing for the first few maturities.
This is actually in stark contrast to the Vasicek model \cite{Hull} 
where the volatility is exponentially decaying with maturity. We will 
see that this universal feature is naturally explained with the 
anticipated trend proposal.

We have studied the {\sc frc} `deformation' determined empirically by:
\begin{equation}
y(t,\theta)=f(t,\theta)-r(t)-\langle s(t,\theta)\rangle_e.
\end{equation}
By construction the deformation process vanishes at $\theta_{\rm min}$ 
and has zero mean. For the first few maturities 
we have observed that this quantity is strongly correlated 
with the past trend in the spot. Therefore, in accordance with the 
anticipated trend proposal, we consider the following simple one-factor 
model:
\begin{equation}
f(t,\theta)=r(t) +\langle s(t,\theta)\rangle +{\cal R}(\theta)b(t).
\end{equation}
The function $b(t)$ is the `anticipated trend' which by 
construction has zero mean. One of the main proposals of \cite{FRC} was that 
the anticipated trend reflects the past trend of the 
spot rate. In other words, the market extrapolates the observed past behaviour of the spot to the 
nearby future. Here we consider a trend of the form:
\be\label{kernel}
b(t)= \int_{-\infty}^t  e^{-\lambda_b(t-t')} d r(t'),
\ee
which corresponds to an exponential cut-off in the past and is equivalent to
an Ornstein-Uhlenbeck process for $b(t)$. 
We have also considered a 
simple flat window cut-off in \cite{frc2}. 
We choose here to calibrate ${\cal R}(\theta)$
to the volatility. Neglecting the contribution of all drifts, we find from
Eq's. (2.6) and (2.7) that the two are related simply by:
\begin{equation}
\sigma(\theta) =\sigma_r\left[1+{\cal R}(\theta)\right].
\end{equation}
In accordance with the observed short-end behaviour of the {\sc frc} volatility, 
we require ${\cal R}(\theta)$ to be {\it positive} and 
strongly increasing for the first few maturities.
In our interpretation of the short-end of the {\sc frc}, 
as described quantitatively by Eq's (2.6-8), this universal feature is a 
consequence of the markets extrapolation of 
the spot trend into the future.

To determine the parameter $\lambda_b$ in Eq. (2.7), we propose to measure 
the following average error: 
\begin{equation}
E=\sqrt{\Bigl< \Bigl(y(t,\theta)-{\cal R}(\theta)b(t)\Bigr)^2\Bigr>}.
\end{equation}
To measure $E$, we must first extract the empirical deformation 
$y(t,\theta)$ using Eq. (2.5). We then determine $b(t)$ using the empirical 
spot time series and Eq. (2.7).\footnote{In this empirical determination of $b(t)$ 
we actually use detrended spot increments, defined as 
$d\hat r(t)=dr(t)-\langle dr \rangle_e $.} 
The error $E$ will have a minimum for some $\lambda_b$. 
This is the time-scale where the deformation and anticipated trend 
match up best, thereby fixing the values of $\lambda_b$. 
\footnote{Note that $E$ is also simply the average error between the empirical 
forward rates and the model forward rates as given by Eq. (2.6).}
In Figure 3 we plot the error $E$, against the 
parameter $\lambda_b^{-1}$, used in the simulation of $b(t)$. 
We consider $\theta=$ 6 months which is the first maturity 
beyond the spot-rate. 
We see a clear minimum demonstrating a strong 
correlation between the deformation and anticipated trend.
For a flat window model the minimum is even more pronounced \cite{frc2}. 
These results indicate the clear presence of a dynamical 
time-scale around $100$ trading days. We have observed 
that the time-scale obtained is independent of the 
maturity used \cite{frc2}.
In Figure 4 we plot the empirical deformation against 
${\cal R}(\theta)b(t)$, where we have 
set $\lambda_b^{-1}=100$ trading days. Indeed, we visually confirm a 
very close correlation. Here we have restricted ourselves 
to a one-factor model for ease of presentation. 
In \cite{frc2} we consider a two and three-factor version 
of our model where the definition of the deformation now includes 
the subtraction of a long spread component. In this case we observe 
improved and very striking correlations that persist even 
up to 2 years forward of the spot! For the other data-sets the strength 
of the correlation is not as strong; however the same qualitative features 
are clearly present.

\section{Comparison with HJM}
It is important to understand whether the popular {\sc hjm} framework 
\cite{hjm} can capture the empirical properties discussed here. 
The stationary one-factor Gaussian {\sc hjm} model is described by:
\begin{equation}
f(t,\theta)=f(t_i,t-t_i+\theta)+\int^t_{t_i}ds~\nu(t+\theta-s)
+\int^t_{t_i}\sigma(t+\theta-s)dW(s),
\end{equation}
where:
\begin{equation}
\nu(\theta)=\sigma(\theta)\int_0^{\theta}d\theta'~\sigma(\theta')
-\lambda\sigma(\theta),
\end{equation}
$\lambda$ is the market price of risk and $dW$ is a Brownian motion. 
The average {\sc frc} is given by:
\begin{equation}
\langle s(t,\theta)\rangle_\tau=f(t_i,\tau+\theta)-f(t_i,\tau)
+\int_{\tau}^{\tau+\theta}du~\nu(u)
-\int_0^{\theta}du~\nu(u),\;\;\tau=t-t_i,
\end{equation}
which corresponds to an average over a finite time period $\tau$. 
For comparisons with our empirical average {\sc frc}, we can consider 
$\tau$ to be 5 years which was the approximate length of our dataset.
There are 3 separate contributions to the average {\sc frc}. 
First is the contribution of the initial {\sc frc}.
In this case the initial {\sc frc} was somewhat steeper than the 
average {\sc frc}. Yet its contribution to the average {\sc frc} 
is still roughly a factor of 3 less than the observed average. 
We can expect the magnitude of this contribution to decrease with 
increasing $\tau$.
The second contribution comes from the $\sigma^2$ factor in Eq. (3.2). 
The magnitude of this contribution grows linearly with $\tau$. Yet even for 
$\tau=10$ years we find that the size of this term is very small, at least a 
factor of 10 smaller than the observed average {\sc frc} for the early 
maturities. This term can therefore be neglected. 
More interesting is the contribution of the 
market price of risk term. We can show that this contribution 
is always {\it negative} for some initial region of the {\sc frc} 
if $\sigma(\theta)>\sigma_r$ for all $\theta$. 
We found that this condition holds for all the data-sets we studied 
\cite{frc2}. This negative contribution has a maximum at 
$\sigma(\theta)=\sigma(\tau+\theta)
\simeq \sigma(\theta_{\max})$. 
Assuming the volatility is constant for large maturities,
we find the market price of risk contribution takes 
the $\tau$ independent form:
\begin{equation}
\langle s(t,\theta)\rangle_{\lambda}
\simeq\lambda\left[\int_0^{\theta}du~\sigma(u)-
\theta \sigma(\theta_{\max})\right].
\end{equation}
In figure 1 we show a plot of 
Eq. (3.4) where we use the empirical volatility Eq. (2.4) and 
choose $\lambda=4.4$ (per $\sqrt{{\rm year}}$) which gives a best fit to the 
average {\sc frc}; it is clear that this fit is very bad, in particular
compared to the simple square-root fit described above. 
In the {\sc usd} case the market price of risk 
contribution is only negative 
for the first maturity since the {\sc usd} has a very strong volatility 
peak. However for the other data-sets it occurs for much longer maturities 
or may remain negative for the entire maturity spectrum. 
Clearly the {\sc hjm} model completely fails to account for 
our empirical results regarding the average {\sc frc}.

The next question to address is whether the {\sc hjm} model 
can explain the striking correlation 
observed between the deformation and anticipated trend. 
We do this by calculating Eq. (2.9), where all averages are 
calculated with respect to the {\sc hjm} model Eq. (3.1) 
calibrated to the empirical volatility. As before we have also 
calibrated ${\cal R}(\theta)$ to the empirical volatility via Eq. (2.8).
An immediate problem arises because, as we have seen, the 
{\sc hjm} average {\sc frc} cannot be calibrated to the empirical 
average {\sc frc}. 
As a result the average deformation will no longer have the required zero mean. 
We will ignore this problem by defining the deformation as Eq. (2.5) 
but with the empirical average {\sc frc} now replaced by the {\sc hjm} 
average {\sc frc}.
In this case we find the finite $\tau$ contributions of Eq. (2.9) are 
negligible and tend to zero for large $\tau$.
The result is plotted in figure 3 where we again consider 
$\theta=6$ months. 
We see that the {\sc hjm} model 
fails to adequately account for the strong anticipated trend effect observed 
here and more strikingly in \cite{frc2}. This is even after we have, 
in effect, assumed that the {\sc hjm} model does describe the correct 
average {\sc frc}. 
On the other hand, our model 
is very close in spirit to the strong correlation limit of the `two-factor' 
spot rate model of Hull-White \cite{hull2}, 
which was introduced in an {\it ad hoc} 
way to reproduce the volatility hump. 
Although phrased differently, this model assumes in effect the existence 
of an anticipated trend following an Ornstein-Uhlenbeck process driven by the 
spot rate \cite{frc2}. It would be interesting to understand better the precise relation, 
if any, between this model and the {\sc hjm} framework \cite{kwon}.

Our main conclusions are as follows. We confirm with much more data that
the average {\sc frc} indeed follows a simple square-root law, with a 
prefactor closely related to the spot volatility. This strengthens the 
idea of a VaR-like pricing of the {\sc frc} proposed in \cite{FRC}. 
We also confirm the striking correlation between the instantaneous 
{\sc frc} and the past spot trend over a certain time horizon. 
This provides a clear empirical confirmation of the anticipated trend 
mechanism first proposed in \cite{FRC}. This mechanism provides a natural 
explanation for the universal qualitative shape of the {\sc frc} volatility at 
the short end of the {\sc frc}. This point is particularly important 
since the short end of the curve is the most liquid part of the curve, 
corresponding to the largest volume of trading (in particular on derivative 
markets). Interest rate models have evolved
towards including more and more factors to account for the dynamics of the
{\sc frc}. Yet our study suggests that after the spot, it is the {\it spot 
trend} which is the most important model component. 
Finally, we saw that the one-factor Gaussian {\sc hjm} model calibrated 
to the empirical 
volatility fails to adequately describe the 
qualitative features discussed here. We presented a simple one-factor 
version of a more complete model described in \cite{frc2}, which is 
consistent with the above interpretation. 

A natural extension of our work is to adapt the general 
method for option pricing in a non-Gaussian world 
detailed in \cite{JPMP} to interest rate derivatives. 
Work in this direction is in progress.

\vskip 0.25cm
{\bf Acknowledgments:}
%\vskip 0.25cm

We thank J. P. Aguilar, P. Cizeau, R. Cont, O. Kwon, L. Laloux, 
M. Meyer, A. Tordjman and in particular M. Potters for many interesting 
discussions.

\begin{figure}[tbp]
\epsfxsize=12cm
\centering{\ \epsfbox{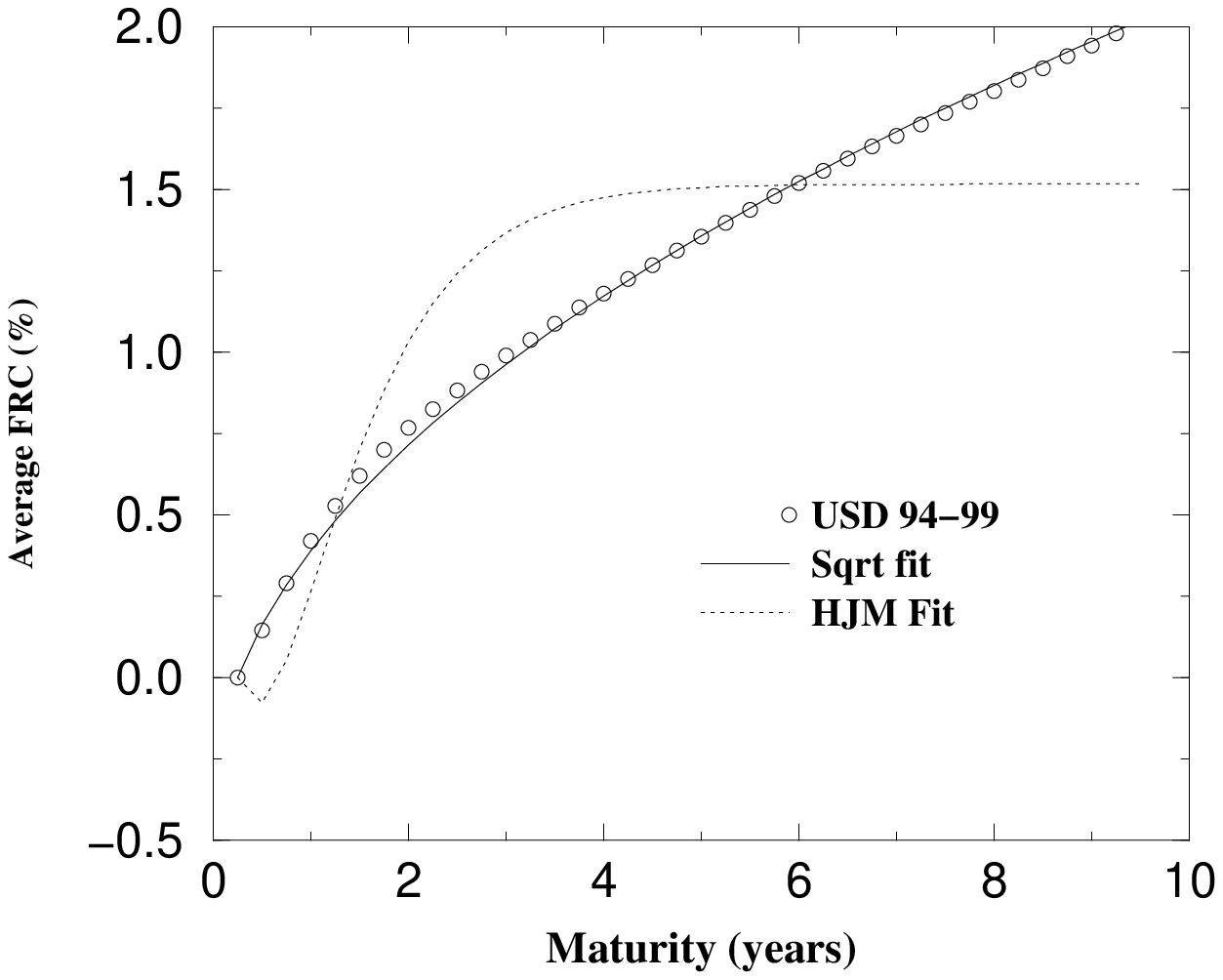}}
\vspace{0.0cm}
\caption{The average {\sc frc} for 
{\sc usd} 94-99, given empirically by Eq. (2.1), and a best 
fit to Eq. (2.2). Also shown is the best fit of Eq. (3.4) which 
is the market price of risk contribution to the average {\sc frc} in the 
{\sc hjm} framework. This figure demonstrates that the {\sc usd} 
average {\sc frc} is  well fitted by a square root law with a prefactor 
given approximately by the spot volatility. The {\sc hjm} framework 
fails to adequately describe this behaviour.}
\end{figure}

\begin{figure}[tbp]
\epsfxsize=14cm
\centering{\ \epsfbox{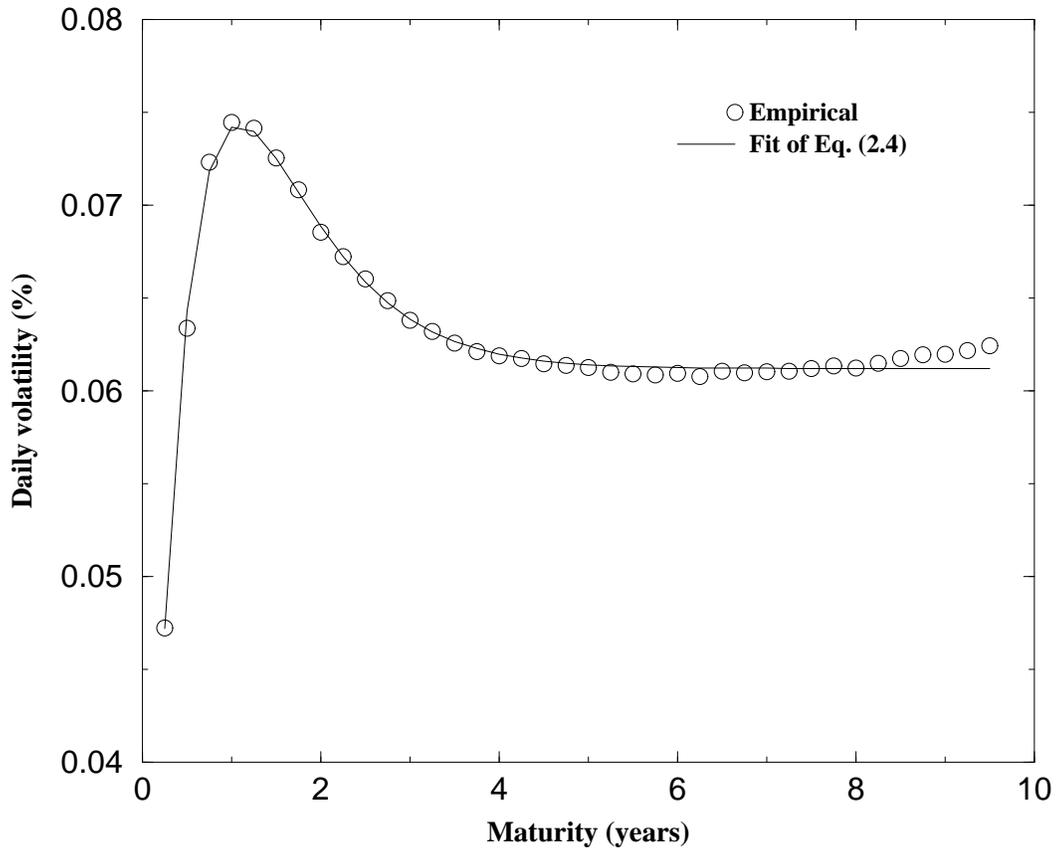}}
\vspace{0.0cm}
\caption{Empirical {\sc frc} volatility for {\sc usd} 94-99 in units of 
\% per square-root day. The empirical volatility is given by Eq. (2.3). 
Also shown is the fit of Eq. (2.4). 
The figure demonstrates that the {\sc frc} volatility has a strong peak 
around a maturity of 1 year.}
\end{figure}

\begin{figure}[tbp]
\epsfxsize=14cm
\centering{\ \epsfbox{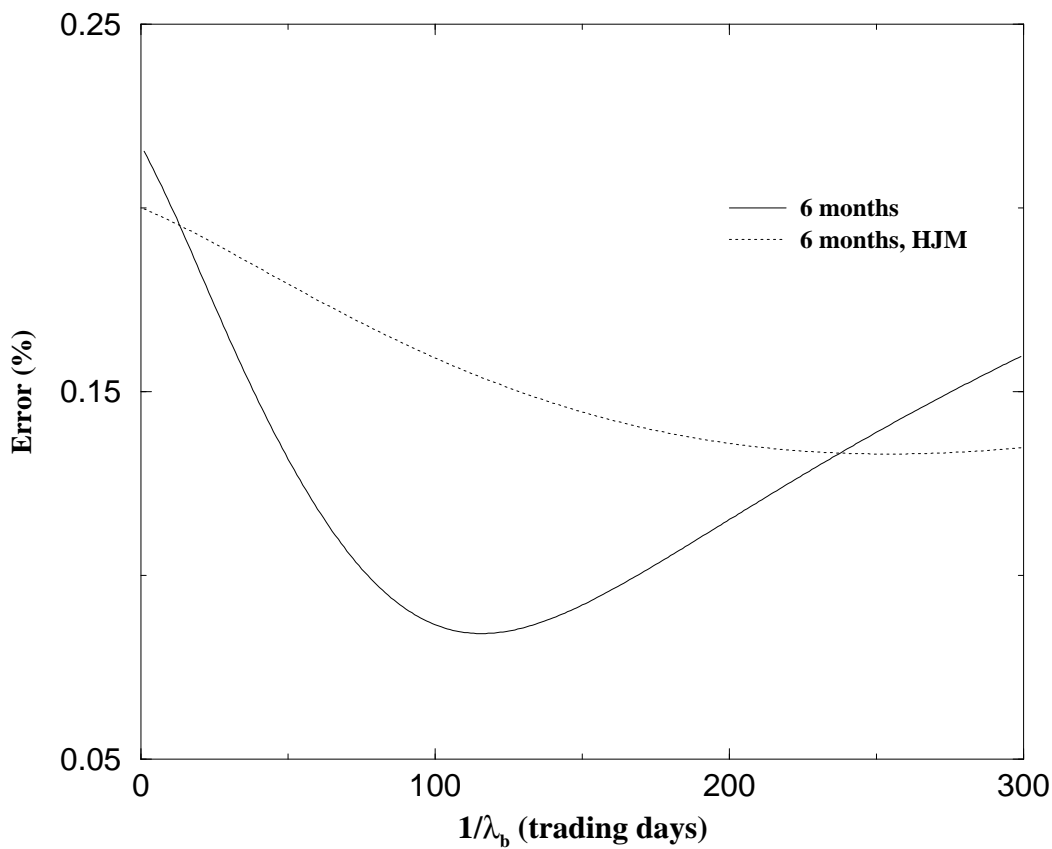}}
\vspace{0.0cm}
\caption{Plot of the error Eq. (2.9), for $\theta=$ 6 months, 
against the parameter $\lambda_b^{-1}$, where for 
the simulation of $b(t)$ we have used Eq. (2.7). Also shown is the error 
predicted by the one-factor Gaussian {\sc hjm} model calibrated to the 
empirical volatility function. 
This figure demonstrates a strong correlation between the deformation 
and anticipated trend along with the clear presence of dynamical time-scale 
in the {\sc usd} {\sc frc}. The {\sc hjm} model fails 
to adequately describe these features.}
\end{figure}

\begin{figure}[tbp]
\epsfxsize=14cm
\centering{\ \epsfbox{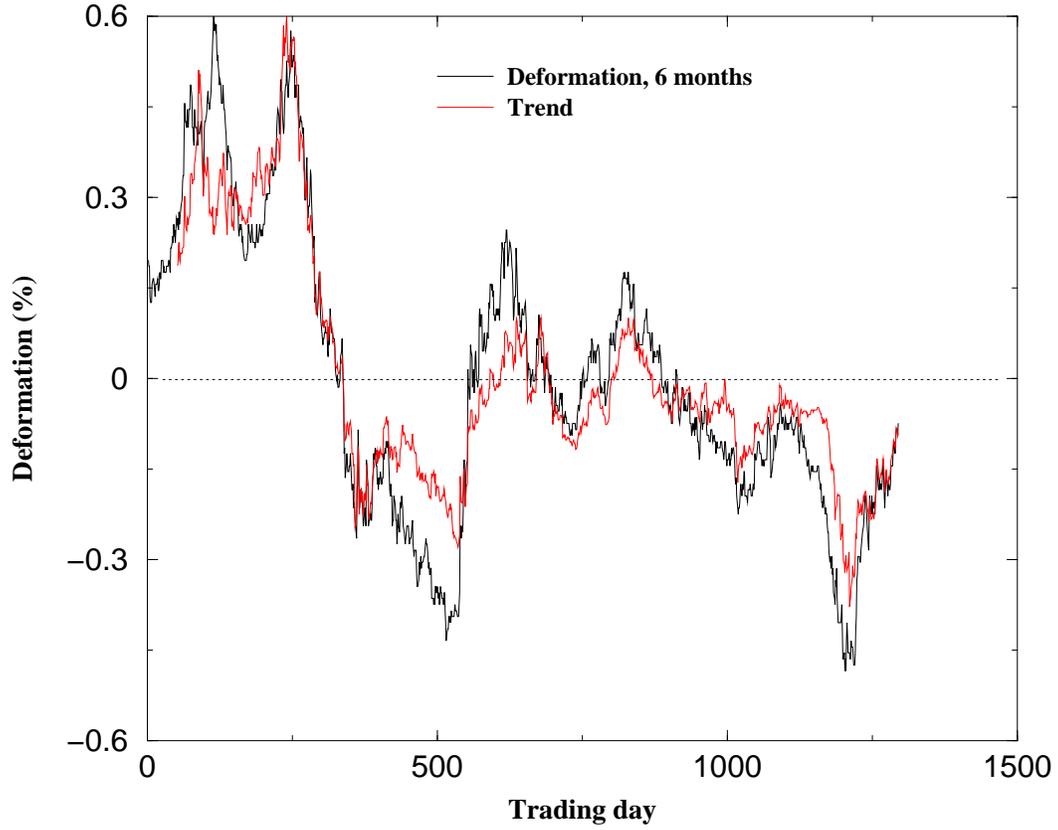}}
\vspace{0.0cm}
\caption{Comparison of the empirical deformation process Eq. (2.5), for 
$\theta=6$ months,
against the scaled anticipated trend, ${\cal R}(\theta)b(t)$. 
We have calculated $b(t)$ using Eq. (2.7), 
with $\lambda_b^{-1}=100$ trading days, while ${\cal R}(\theta)$ is 
calibrated to the {\sc frc} volatility. 
The period covered is 1/1/94 to 18/2/99.
This figure demonstrates a strong correlation 
between the empirical deformation process and the anticipated trend.}
\end{figure}

\end{document}